\documentclass{article}
\usepackage{epsfig}
\newcommand{\bfr}{\begin{flushright}}
\newcommand{\efr}{\end{flushright}}
 
\begin{document}
\title{Classical and Quantum Scattering of Maximally Charged Dilaton
Black Holes
}
\author{Kiyoshi Shiraishi\\
Akita Junior College, Shimokitade-Sakura,\\
Akita-shi, Akita 010,
Japan}
\date{International Journal of Modern Physics \textbf{D2} (1993)
pp.~59--77}
\maketitle
\begin{abstract}
The classical and the quantum scattering of two maximally-charged
dilaton black holes which have low velocities are studied. We find a
critical value for the dilaton coupling, $a^2=1/3$. For $a^2>1/3$, two
black holes are always scattered away and never coalesce together,
regardless of the value of the impact parameter.
\end{abstract}

\section{Introduction}
The slow motion of solitons is described by a geodesic motion in a
finite-dimensional moduli space if several forces between the static
solitons are balanced with one another or are assumed to be very weak.

An example is found in the system consisting of BPS monopoles. This
system has been extensively studied for a decade \cite{1,2,3,4,5,6}.
The magnetic repulsive force and Higgs scalar force between monopoles at
rest are cancelled by each other in this case. The slow motions of BPS
monopoles are analyzed by using the metric on the moduli space
\cite{1,2,3,4,5,6}.

Another example is found in the multi-vortex system in the Abelian
Higgs model with a critical relation between gauge and Higgs couplings.
In this system, the electromagnetic force and scalar force are not only
reduced to short-range forces but also cancelled by each other. The
vortex moduli space is also investigated by several authors
\cite{7,8,9,10}.

In this paper, we examine a special kind of multi-black-hole system. Originally, 
dynamics of extreme Reissner--Nordstrom (RN) black holes at low energy
was con-  sidered by Gibbons and Ruback \cite{12}. In this system, the
strength of the gravitational  force is equal to that of the Coulomb
force between the extreme RN black holes.  The moduli space metric for
the extreme RN black holes at arbitrary distances was  shown by Ferrell
and Eardley \cite{13}. More recently, Traschen and Ferrell \cite{14}
investigated  quantum-mechanical scattering of the extremal RN holes by
quantizing the moduli.  

The charged black holes in the system we study
here are assumed to be coupled  to dilaton fields
\cite{15,16,17,18,19,20,21,22,23}. Thus a (attractive) scalar force
is involved in the system.  The multi-centered static configuration of
maximally charged dilaton black holes was shown by the present author
recently \cite{22}. In this case, three forces are balanced with one
another in the static system described by the solution. The metric on
the moduli space was also obtained in Ref.~\cite{23}. In this paper, the
scattering of two maximally-charged dilaton black holes is more closely
investigated by classical and quantum-mechanical methods.

The dynamics of the extremal holes may depend on the dilaton coupling.
The existence of critical values for the coupling has recently been
pointed out by several authors through the study of the classical and
quantum nature of the charged dilaton black holes
\cite{15,16,17,18,19,20,21,22,23}. We will rediscover the critical
behavior in the present study on classical and quantum treatments of the
moduli space.

In Sec.~2, the static solution in the Einstein-Maxwell-dilaton system
is reviewed and the derivation of the metric on moduli space is
discussed.

In Sec.~3, the classical two-body scattering is studied. We compare the
result of a small-particle approximation with that of a geodesic
approximation on the moduli space in the corresponding limit. The
comparison is absolutely necessary to confirm the validity of the
approximation, even if it does not provide much knowledge. The
cross-section is found, especially for the case with the dilaton
coupling of string theory.

In Sec.~4, the quantum-mechanical scattering is examined. Again, we
compare the wave scattering by a black hole and the description by a
quantization procedure on the collective motion of black holes. The
cross-section for partial waves are discussed.

The conclusion is given in Sec.~5. Throughout this paper, we deal with
a $(1+3)$-dimensional spacetime.

\section{The Moduli Space Metric for the System Consisting
of Maximally Charged Dilaton Black Holes}
The Einstein-Maxwell-dilaton system contains a dilaton field $\phi$
coupled to a $U(1)$ gauge field $A_\mu$, besides the Einstein-Hilbert
gravity.

The action for the fields with particle sources is
\begin{equation}
S=\int
d^4x\frac{\sqrt{-g}}{16\pi}[R-2(\nabla\phi)^2-e^{-2a\phi}
F_{\mu\nu}F^{\mu\nu}]-\sum_a\int
ds_a\left[m_ae^{a\phi}+e_a A_\mu\frac{dx_a^\mu}{ds_a}\right]\,,
\label{2.1}
\end{equation}
where $R$ is the scalar curvature and $F_{\mu\nu}=\partial_\mu A_\nu-
\partial_\nu A_\mu$.

The Newton constant has been set to be unity. The dilaton coupling to
the Maxwell term is governed by a constant $a$. Here the coupling to the
point source has already been fixed for later use. $s_a$ parametrizes
each worldline of the particle.

The metric for the $N$-body system of maximally-charged dilaton black
holes has been known as \cite{22}
\begin{equation}
ds^2=-U^{-2}(x^k)dt^2+U^2(x^k)\delta_{ij}dx^idx^j\,,
\label{2.2}
\end{equation}
with
\begin{equation}
U(x^k)=\{F(x^k)\}^{1/(1+a^2)}
\label{2.3}
\end{equation}
and
\begin{equation}
F(x^k)=1+\sum_{a=1}^N\frac{(1+a^2)m_a}{|\mathbf{x}-\mathbf{x}_a|}\,.
\label{2.4}
\end{equation}

Using these expressions, the vector one-form and dilaton configuration
are written as
\begin{equation}
A=\left(\frac{1}{1+a^2}\right)^{1/2}[1-\{F(x^k)\}^{-1}] dt
\label{2.5}
\end{equation}
and
\begin{equation}
e^{-2a\phi}=\{F(x^k)\}^{2a^2/(1+a^2)}\,.
\label{2.6}
\end{equation}
In this solution, the asymptotic values of $\phi$ and $A$ are fixed to be
zero.

The electric charge $e_a$ and scalar charge $\sigma_a$ of each black hole
are associated with the corresponding mass $m_a$ by
\begin{eqnarray}
e_a&=&(1+a^2)^{1/2}m_a\,,
\label{2.7}\\
\sigma_a&=&a m_a\,.
\label{2.8}
\end{eqnarray}

One can explicitly verify cancellation of static forces between the
black holes at large distances using the relations. If we set $a=0$ in
the solution, we find that the solution reduces to the
Papapetrou-Majumdar solution \cite{24}. The dimension of the parameter
space of the static solution is $3N$, which is the number of coordinates
describing the position of $N$ black holes.

We anticipate that an addition of a small amount of kinetic energy to
this static system can be treated by perturbation. This is justified on
the assumption that radiation reactions can be ignored at low velocity.

Although we can obtain the Lienard-Wiechert potential at large
distances in the system \cite{12}, we must apply the method of Ferrell
and Eardley \cite{13} to the system in order to get the information on
the slow motion of the extremal black holes at arbitrary distances.

The perturbed metric and potential can be written in the form
\begin{eqnarray}
ds^2&=&-U^{-2}dt^2+2N^idx^idt+U^2dx^idx^i\,,
\label{2.9}\\
A&=&\left(\frac{1}{1+a^2}\right)^{1/2}\{1-[F(x^k)]^{-1}\}dt+A_idx^i\,,
\label{2.10}
\end{eqnarray}
where $U$ and $F(x)$ are defined by (\ref{2.3}) and (\ref{2.4}). We need
only to solve linearized equations with perturbed sources [where
$e_a=(1+a^2)^{1/2}m_a$ is substituted] up to $O(v)$ for $N_i$ and $A_i$
for our purpose. Each source plays the role of a maximally charged
dilaton black hole.

The authors of Refs.~\cite{13} and \cite{14} have emphasized the
``smearing'' of the source to  avoid the violet divergences. It is known
that divergences often appear in gravitating systems \cite{25,26}.
The systematic regularization procedure is rather technical and yields
no ambiguity.

Solving the Einstein and Maxwell equations and substituting the
solutions, the perturbed dilaton field and sources to the
field-theoretical action (\ref{2.1}) with proper boundary terms, we get
the effective Lagrangian up to $O(v^2)$ for the
$N$-maximallyÊcharged-dilaton-black-hole system: \cite{23}
\begin{eqnarray}
L&=&-\sum_a m_a+\sum_a\frac{1}{2}m_av_a^2\nonumber \\
&=&\frac{3-a^2}{8\pi}\int d^3x\,[F(x)]^{2(1-a^2)/(1+a^2)}\sum_{a,b}
\frac{(\mathbf{n}_a\cdot\mathbf{n}_b)(\mathbf{v}_a-\mathbf{v}_b)^2
m_am_b}{2|\mathbf{r}_a|^2|\mathbf{r}_b|^2} \,,
\label{2.11}
\end{eqnarray}
where $\mathbf{r}_a=\mathbf{x}-\mathbf{x}_a$ and $\mathbf{n}_a=
\mathbf{r}_a/|\mathbf{r}_a|$. $F(x)$ is defined by (\ref{2.4}).

Sending the value of dilaton coupling $a$ to zero in the above formal
expression, we reproduce the result of Ref.~\cite{13}. Furthermore, in
the large-distance limit [where we approximate $F(x)$ by 1], we recover
exactly the same result obtained by the Lienard-Wiechert method
\cite{12}.

Generally there exist many-body interactions. In general, we obtain
infinite species of many-body interactions by expanding the function
$F(x)$. The following are some special cases:

\begin{itemize}
\item	$a^2=0$: The black holes are governed only by two-body, three-body
and four-body interactions. \cite{13}
\item	$a^2=1/3$: There are two-body and three-body interactions.
\item	$a^2=1$: There are only two-body interactions.
\item	$a^2=3$: There is no interaction.
\end{itemize}

The general expression (\ref{2.11}) needs further regularization on
performing the integration. If one wants to study the slow motion of the
maximally-charged dilaton black holes in the low-velocity limit, one has
to analyze a two-body system at the first step. We consider a
two-black-hole system hereafter.

After regularization, the effective Lagrangian for a two-body system
(consisting of black holes labelled with $a$ and $b$) can be rewritten as
\begin{eqnarray}
L_{2B}&=&-M+\frac{1}{2}MV^2\nonumber \\
& &+\frac{1}{2}\mu v^2\left\{1-\frac{M}{\mu}-\frac{(3-a^2)M}{r}+
\frac{M}{m_a}\left[1+\frac{(1+a^2)m_a}{r}\right]^{(3-a^2)/(1+a^2)}
\right.\nonumber \\
&
&+\left.\frac{M}{m_b}
\left[1+\frac{(1+a^2)m_a}{r}\right]^{(3-a^2)/(1+a^2)}\right\}\,,
\label{2.12}
\end{eqnarray}
where $M=m_a+m_b$, $\mu=m_am_b/M$,
$\mathbf{V}=(m_a\mathbf{v}_a+m_b\mathbf{v}_b)/M$,
$\mathbf{v}=\mathbf{v}_a-\mathbf{v}_b$ and
$r=|\mathbf{x}_a-\mathbf{x}_b|$.

Since the motion of the center-of-mass is separable, we focus on the
relative motion even if we proceed to the quantum-mechanical two-body
problem. In terms of the metric on the moduli space, the separation of
the motion means that the metric takes the block-diagonal form.

The metric for the relative motion is
\begin{equation}
g_{ab}=\gamma(r)\delta_{ab}\,,
\label{2.13}
\end{equation}
with
\begin{eqnarray}
\gamma(r)&=&1-\frac{M}{\mu}-\frac{(3-a^2)M}{r}+
\frac{M}{m_a}\left[1+\frac{(1+a^2)m_a}{r}\right]^{(3-a^2)/(1+a^2)}
\nonumber \\
&
&+frac{M}{m_b}
\left[1+\frac{(1+a^2)m_a}{r}\right]^{(3-a^2)/(1+a^2)}\,.
\label{2.14}
\end{eqnarray}

In particular, when $m_a=m_b=m$ (then $M=2m$ and $\mu=m/2$), $\gamma(r)$
becomes
\begin{equation}
\gamma(r)=1-\frac{(3-a^2)M}{r}+
4\left[1+\frac{(1+a^2)M}{2r}\right]^{(3-a^2)/(1+a^2)}-4\,.
\label{2.15}
\end{equation}

On the other hand, if $m_b\ll m_a$ (then $M\sim m_a$), i.e.,
$\mu\rightarrow 0$, we obtain
\begin{equation}
\gamma(r)=\left[1+\frac{(1+a^2)M}{2r}\right]^{(3-a^2)/(1+a^2)}\,.
\label{2.16}
\end{equation}

We examine the cases with some special values for $a$.

For $a=0$, we recover the result of Ref.~\cite{13}
\begin{equation}
\gamma(r)=1+\frac{3M}{r}+\frac{3M^2}{r^2}+\frac{M^3}{r^3}
\left(1-\frac{2\mu}{M}\right)\,.
\label{2.17}
\end{equation}
(Ref.~\cite{14} includes misprints at this point.)

For $a^2=1/3$, we find
\begin{equation}
\gamma(r)=\left(1+\frac{4M}{3r}\right)^2\,.
\label{2.18}
\end{equation}

For $a^2=1$, the value corresponding	to the string theory, we find 
\begin{equation}
\gamma(r)=1+\frac{2M}{r}\,.
\label{2.19}
\end{equation}
Note that $\gamma(r)$ does not depend on the reduced mass $\mu$ for both
cases $a^2=1/3$ and $a^2=1$.

For $a^2=3$, we get an obvious result
\begin{equation}
\gamma(r)=1\,.	
\label{2.20}
\end{equation}

In the following section, we will describe the classical scattering of
two maximally charged dilaton black holes using the metric on the moduli
space

\section{The Classical Scattering of Two Maximally
Charged Dilaton Black Holes}
Once the metric is known, it is easy to compute geodesic motion on the
moduli space. We pay attention to the two-body scattering problem. We
assume that the scattering occurs on a scattering plane, where the
distance $r$ and the azimuthal angle $\theta$ determine the
configuration. Then the metric on the moduli space for the two-body
system is reduced to
\begin{equation}
ds_{MS}^2=\gamma(r)(dr^2+r^2d\theta^2)\,.
\label{3.1}
\end{equation}
To analyze the geodesic trajectory on the plane is straightforward. We
find that the trajectory is described by the differential equation
\begin{equation}
\left(\frac{dr}{d\theta}\right)^2=r^4\left[
\frac{\gamma(r)}{b^2}-\frac{1}{r^2}\right]\,,
\label{3.2}
\end{equation}
where $b$ denotes the impact parameter.

If the mass of one of the black holes, $M$, overwhelms the other, the
equation becomes
\begin{equation}
\left(\frac{dr}{d\theta}\right)^2=r^4\left\{
\frac{1}{b^2}\left[1+\frac{(1+a^2)M}{r}
\right]^{(3-a^2)/(1+a^2)}-\frac{1}{r^2}\right\}\,.
\label{3.3}
\end{equation}

Before solving the equation, we consider the point-particle equations
of motion in	the background field of a maximally charged dilaton
black hole with mass $M$. In this
case, we can identify the coordinates of background geometry with the
collective coordinates on the moduli space. The particle, playing
the role of a maximally charged dilaton black hole, has mass $= m$
and charge $=e=(1+a^2)^{1/2}m$. The action for the particle is
\begin{equation}
\int ds \left[m e^{a\phi} +eA_\mu\frac{dx^\mu}{ds}\right]\,,	
\label{3.4}
\end{equation}
[taking the same form as in (\ref{2.11})] and the background metric and
fields are expressed by (\ref{2.2}, \ref{2.3}, \ref{2.5}, \ref{2.6})
with
\begin{equation}
F(r)=1+\frac{(1+a^2)M}{r}\,.
\label{3.5}
\end{equation}

After some manipulation, we can show that the equation of motion in the
background fields is once integrated to become
\begin{equation}
\left(\frac{dr}{d\theta}\right)^2=r^4\left[
\left(\frac{E-m}{L}\right)^2F^{4/(1+a^2)}+
\frac{2m(E-m)}{L^2}F^{(3-a^2)/(1+a^2)}
-\frac{1}{r^2}\right]\,,
\label{3.6}
\end{equation}
where the constants $E$ and $L$ correspond respectively to energy and
angular momentum of the incident particle at spatial infinity. The
energy and angular momentum at infinity can be written as
$E=m(1-v^2)^{-1/2}$ and $L=mvb(1-v^2)^{-1/2}$, where $v$ is the
asymptotic velocity of the incident particle. Then (\ref{3.6}) becomes
\begin{equation}
\frac{1}{r^4}\left(\frac{dr}{d\theta}\right)^2=
\frac{1}{b^2}F^{(3-a^2)/(1+a^2)}-\frac{1}{r^2}+
\frac{v^2}{4b^2}F^{4/(1+a^2)}+O(v^4)
\,.
\label{3.7}
\end{equation}

Apparently, in the low-velocity limit $v\rightarrow	0$, the result
(\ref{3.7}) coincides with the previous result (\ref{3.3}) (for any
$a^2$) which came from analyzing the moduli metric. Note that the moduli
space metric in a general case contains more information; the geodesic
approximation can quantitatively describe the motion in the system with
arbitrary $m$ and $\mu$ in the low-velocity limit.

Now we study the two-body problem in the low-velocity limit. First, we
examine the classical condition that the two extremal dilaton black
holes coalesce. In the geodesic approximation, two black holes coalesce
if the following algebraic equation for $r$ has no positive root:
\begin{equation}
r^2\gamma(r)=b^2\,.
\label{3.8}
\end{equation}
It is possible that the equation has a positive solution for any value
of $b$. In such a case, black holes are always scattered away (provided
that the low-velocity approximation is valid).

Next, we examine the deflection angle $\Theta$. $\Theta=\pi$ corresponds
to the forward scattering, while $\Theta=0$ corresponds to the backward
scattering. Since the particle (black hole) is scattered away if
(\ref{3.8}) has a solution, the deflection angle is expressed in the
integral form
\begin{equation}
\Theta=\int_{r_0}^\infty \frac{2b\,dr}{r\sqrt{r^2\gamma(r)-b^2}}-\pi\,,
\label{3.9}
\end{equation}
where $r_0$ is the maximum positive solution for Eq.~(\ref{3.8}).

Then, the classical differential cross-section (per solid angle) can be
computed:
\begin{equation}
\frac{d\sigma}{d\Omega}=\left|\frac{\sin\Theta}{b}
\frac{d\Theta}{db}\right|^{-1}\,.
\label{3.10}
\end{equation}

The above expressions are valid for general cases, i.e., the collective
motion of maximally-charged dilaton black holes. For a while, we focus
our attention on a particular case where the mass of the ``target''
black hole ($M$) is much larger than the ``incident'' black hole. In this
case, the reduced mass $\mu$ becomes infinitesimal. Then, we must solve
(\ref{3.3}), or apply Eqs.~(\ref{3.8})-(\ref{3.10}) to the case with the
moduli space metric (\ref{2.16}).

Considering the coalescence condition, we distinguish the following
cases.

\begin{itemize}
\item	$a^2\le 1/3$: There is a critical value $b_{coal}$ of $b$, inside
of which black holes coalesce \cite{13}. $b_{coal}$ turns out to be
\begin{eqnarray}
b_{coal}&=&\frac{1-3a^2}{2}\left(\frac{3-a^2}{1-3a^2}
\right)^{(3-a^2)/[2(1+a^2)]}M\quad\mbox{for } a^2<1/3\,,
\label{3.11}\\
b_{coal}&=&(4/3)M\qquad\qquad\qquad\qquad\qquad\quad
\quad\quad~~\mbox{for }
a^2=1/3\,.
\label{3.12}
\end{eqnarray}
\item	$a^2>1/3$: In this case, black holes never coalesce for any value
of $b$ (provided that the slow-velocity approximation is used).
\end{itemize}

The scattering cross-section can be analytically obtained only for
special values of $a^2$. The most impressive and simple result is found
in the case with $a^2=1$, i.e., the coupling which appears in string
theory. The deflection angle is given by
\begin{equation}
\Theta=2\tan^{-1}\frac{M}{b}\qquad (a^2=1)\,,
\label{3.13}
\end{equation}
and then, the classical differential cross-section is
\begin{equation}
\frac{d\sigma}{d\Omega}=\frac{1}{4}
\frac{M^2}{\sin^4(\Theta/2)}\qquad (a^2=1)\,.
\label{3.14}
\end{equation}

This has precisely the same $\Theta$-dependence as the Rutherford
scattering! In this case, the deflection angle $\Theta$ approaches $\pi$
as $b/M$ goes to zero. This behavior can be attributed to the deficit
angle near the origin on the moduli space for $a^2=1$ \cite{23}. Note
that, in this case with $a^2=1$, the result (\ref{3.13}) and (\ref{3.14})
holds for any masses of black holes provided that $M=m_a+m_b$.

We can relate the qualitative features of scattering (or coalescence) to
the shape of the moduli space for the two-body system with arbitrary
$a^2$ and masses. The metric on the (reduced) moduli space (\ref{3.1})
can be rewritten by coordinate transformation as
\begin{equation}
ds_{MS}^2=h(R)dR^2+R^2d\theta^2\,,
\label{3.15}
\end{equation}
where $R=\gamma^{1/2}r$ and
\begin{equation}
h(R(r))=\left(1+\frac{r}{2\gamma}\frac{d\gamma}{dr}\right)^{-2}\,.
\label{3.16}
\end{equation}

For $a^2<1/3$, $R$ increases as the separation $r$ goes to zero; thus the
surface of the moduli space immersed into three dimensions seems to have
a ``throat'' \cite{13,14}. If the trajectory goes across the throat, it
represents the coalescence of black holes. For $a^2= /3$, $R$ approaches
a constant when $r$ is reduced to zero. The surface looks like a funnel.
For $a^2>1/3$, both $R$ and $r$ approach zero at the same time. In the
vicinity of the origin, $R=0$, the metric is approximated by
\begin{equation}
ds_{MS}^2=\left[\frac{2(1+a^2)}{3a^2-1}\right]^{2}dR^2+R^2d\theta^2
\qquad (R\sim 0, \mbox{for } a^2>1/3)\,.
\label{3.17}
\end{equation}
One can show that there is a deficit angle $\Delta\theta$ around the
origin:
\begin{equation}
\Delta\theta=\frac{3-a^2}{1+a^2}\pi
\qquad (R\sim 0, \mbox{for } a^2>1/3)\,.
\label{3.18}
\end{equation}
This result is independent of the masses of black holes. Therefore, for
$a^2>1/3$, the deflection angle $\Theta$ approaches $\Delta\theta$ in the
limit of small $b$.

\section{The Quantum Scattering of Two Maximally
Charged Dilaton Black Holes}
In the previous section, we have seen that the classical low-energy
scattering of two maximally charged dilaton black holes is well
described by a geodesic motion on the (reduced) moduli space. We
consider the quantum scattering in this section. The quantization of the
moduli parameters has been discussed in Refs.~\cite{3,4,11,14} for
various systems.

We first review the quantization on the moduli space. Let us introduce a
wave function $\Psi$ on the moduli space, which obeys the Schrodinger
equation \cite{14}
\begin{equation}
i\hbar\frac{\partial\Psi}{\partial t}=\left(
-\frac{\hbar^2}{2\mu}\nabla^2+\hbar^2\xi R_{(MS)}\right)\Psi\,,
\label{4.1}
\end{equation}
where $\nabla^2$ is the covariant Laplacian constructed from the moduli
space metric and $R_{(MS)}$ is the scalar curvature of the moduli space.
We assume $\xi=0$ in this paper though this term may be present in most
general cases. Here, we have already removed the center-of-mass degrees
of freedom, as usual.

The partial wave in a stationary state is
\begin{equation}
\Psi=\psi_{ql}(r) Y_{lm}(\theta, \phi) \exp(-iEt/\hbar)\,,
\label{4.2}
\end{equation}
where $Y_{lm}(\theta, \phi)$ is the spherical harmonic function and $E=
\hbar^2q^2/(2\mu)$.

Recalling the metric of the form (\ref{2.13}), we can rewrite the
equation for the partial wave function as \cite{14}
\begin{equation}
\psi''+\frac{2}{r}\psi'+\frac{\gamma'}{2\gamma}\psi'-
\frac{l(l+1)}{r^2}\psi+q^2\gamma\psi=0\,,
\label{4.3}
\end{equation}
where the prime denotes the derivative with respect to $r$.

Now, we compare the Schrodinger equation with the scalar wave function,
whose quantum has small mass m and charge $(1+a^2)^{1/2}m$, in the
background metric and fields of a maximally charged dilaton black hole
with mass $=M$. The scalar field is obtained from the second
quantization of the particle action (\ref{3.4}). The wave function is
given by
\begin{equation}
\hbar^2(\nabla_\mu+iA_\mu)e^{-2ab\phi}(\nabla^\mu+iA^\mu)\psi
-e^{-2ac\phi}m^2\psi= 0\,,
\label{4.4}
\end{equation}
where the constants $b$ and $c$, which satisfy $b-c=1$ \cite{19}, arise
from an ambiguity in  the procedure of the second quantization. The
partial wave decomposition leads to
\begin{equation}
\Psi=\psi_{ql}(r) Y_{lm}(\theta, \phi) \exp(-i\omega t)\,,
\label{4.5}
\end{equation}
where $\hbar\omega=(m^2+\hbar^2q^2)^{1/2}$.

Substituting (\ref{2.2}), (\ref{2.3}), (\ref{2.5}) and (\ref{2.6}) with
(\ref{3.5}) into (\ref{4.4}), we obtain the differential equation for the
partial wave:
\begin{eqnarray}
& &\psi''+\frac{2}{r}\psi'+\frac{X'}{2X}\psi'-
\frac{l(l+1)}{r^2}\psi\nonumber \\
& &\qquad+F^{4/(1+a^2)}\frac{1}{\hbar^2}
\left[(\hbar\omega-m)^2+\frac{2m(\hbar\omega-m)}{F}\right]\psi=0\,,
\label{4.6}
\end{eqnarray}
with $X=e^{-4ab\phi}$.

In the nonrelativistic and low-energy limit, $\hbar q\ll m$, the wave
equation takes the following form:
\begin{equation}
\psi''+\frac{2}{r}\psi'+\frac{X'}{2X}\psi'-
\frac{l(l+1)}{r^2}\psi
=-F^{(3-a^2)/(1+a^2)}\frac{1}{\hbar^2}
q^2\psi+O(q^4)\,.
\label{4.7}
\end{equation}

This coincides with Eq.~(\ref{4.3}), up to the term originating from an
ambiguity of ``operator ordering''. The difference coming from two
approaches is expected to be unimportant provided that $\psi'$ is small
near $r\sim 0$.

Let us analyze the scattering problem, adopting the wave equation on the
moduli space (\ref{4.3}). (Hereafter, we set $\hbar=1$ unless otherwise
stated, for simplicity.) In this paper, we do not intend to survey the
scattering problem for all possible values of $a^2$. We pick up typical
cases and use some typical approximation method to show peculiarities of
charged dilaton black holes.

The differential equation can be written in a simple form by introducing
a new 	coordinate $R=\int\sqrt{\gamma}	dr$ and writing $\psi=
\chi/(r\sqrt{\gamma})$. Then, one can find \cite{14}
\begin{equation}
\chi,_{RR}+(q^2-V)\chi=0\,,
\label{4.8}
\end{equation}
with
\begin{eqnarray}
V&=&\frac{1}{2r\gamma}\left(\frac{r\gamma'}{\gamma}
\right)'+\frac{l(l+1)}{r^2\gamma}\nonumber \\
&=&\frac{\gamma(r\gamma')'-r(\gamma')^2}{2r\gamma^3}+
\frac{l(l+1)}{r^2\gamma}\,.
\label{4.9}
\end{eqnarray}
(Ref.~\cite{14} includes misprints at this point.)

As in the previous section, we treat the case with $\mu\rightarrow 0$,
i.e., $m_b\ll m_a\sim M$. We have to remark that the results hold good
for the cases with arbitrary values of $\mu$ if $a^2$ takes some special
values such as $1$ or $1/3$.

In the limit of $\mu\rightarrow 0$, $V$ takes the form
\begin{eqnarray}
V(r(R))&=&\frac{(3-a^2)\left[\frac{r}{(1+a^2)M}
\right]^{2(1-a^2)/(1+a^2)}}{2(1+a^2)^3M^2\left[1+\frac{r}{(1+a^2)M}
\right]^{(5+a^2)/(1+a^2)}}\nonumber
\\ & &+
\frac{l(l+1)}{(1+a^2)^2M^2}\frac{\left[\frac{r}{(1+a^2)M}
\right]^{(1-3a^2)/(1+a^2)}}{\left[1+\frac{r}{(1+a^2)M}
\right]^{(3-a^2)/(1+a^2)}}\,.
\label{4.10}
\end{eqnarray}

The range of the new coordinate $R$ depends on the dilaton coupling
$a^2$. For $a^2\le 1/3$, $R$ varies from $-\infty$ to $+\infty$, as $r$
varies from $0$ to $+\infty$. For $a^2<1/3$,
\[
R\sim r+\left(\frac{3-a^2}{2}\right)M\ln\left[
\frac{r}{(1+a^2)M}\right]\quad\mbox{as }
r\rightarrow\infty\quad \mbox{and}
\]
\[
R\sim -\left[\frac{2(1+a^2)^2}{1-3a^2}\right]M\left[
\frac{r}{(1+a^2)M}\right]^{-(1-3a^2)/[2(1+a^2)]}\quad\mbox{as }
r\rightarrow 0\,.
\]
For $a^2= 1/3$, $R=r+(4/3)M\ln(3r/4M)$.

For $a^2>1/3$, the range of $R$ is $[0, +\infty]$. Asymptotically,
\[
R\sim r+\left(\frac{3-a^2}{2}\right)M\ln\left[
\frac{r}{(1+a^2)M}\right]\quad\mbox{as }
r\rightarrow\infty\quad \mbox{and}
\]
\[
R\sim \left[\frac{2(1+a^2)^2}{3a^2-1}\right]M\left[
\frac{r}{(1+a^2)M}\right]^{(3a^2-1)/[2(1+a^2)]}\quad\mbox{as }
r\rightarrow 0\,.
\]

Consequently, we can draw the shape of the potential $V(R)$. In the
asymptotic region, we find
\begin{equation}
V(R)\sim\frac{(3-a^2)M}{2R^3}+\frac{l(l+1)}{R^2}
\quad\mbox{for } R/(1+a^2)M\gg 1 ~(r\rightarrow\infty)\,.	
\label{4.11}
\end{equation}

At the small separation length, on the other hand, $V(R)$ looks like
\begin{equation}
V(R)\sim\frac{(3-a^2)}{2(1+a^2)^3M^2}\left[
\frac{(3a^2-1)R}{2(1+a^2)M}
\right]^{-4(1-a^2)/(1-3a^2)}
+\frac{4(1+a^2)^2}{(1-3a^2)^2}\frac{l(l+1)}{R^2}\,.	
\label{4.12}
\end{equation}
This should read as $V(R\sim-\infty)$ for $a^2<1/3$, while as $V(R\sim
0)$ for $a^2>1/3$. 

For the critical value $a^2=1/3$, as $R\rightarrow\infty$, the
potential becomes
\begin{equation}
V(R)=\frac{9}{16M^2}\exp\left(
\frac{3R}{4M}\right)
+\frac{9l(l+1)}{16R^2}\exp\left(-\frac{3R}{2M}\right)
\quad\mbox{for } R\rightarrow-\infty\,.	
\label{4.13}
\end{equation}

\begin{figure}[ht]
\begin{center}
\includegraphics[width=5cm]{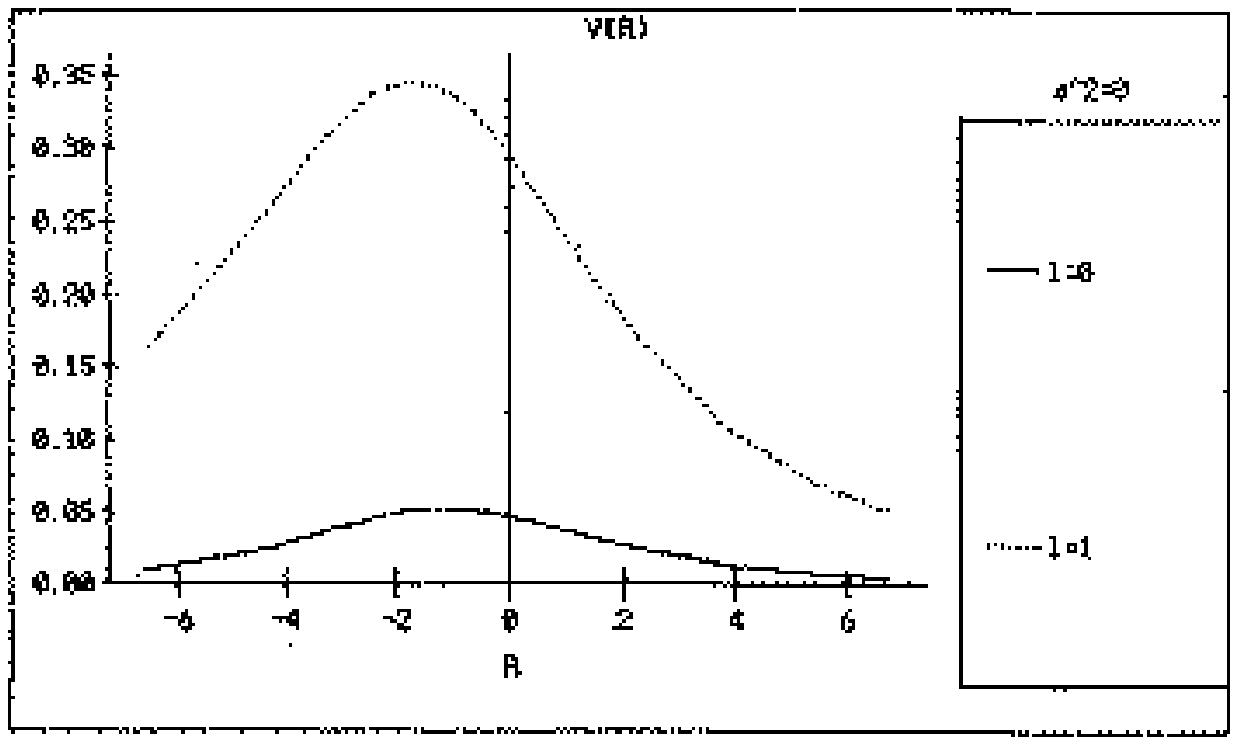}\\
(a)\\
\includegraphics[width=5cm]{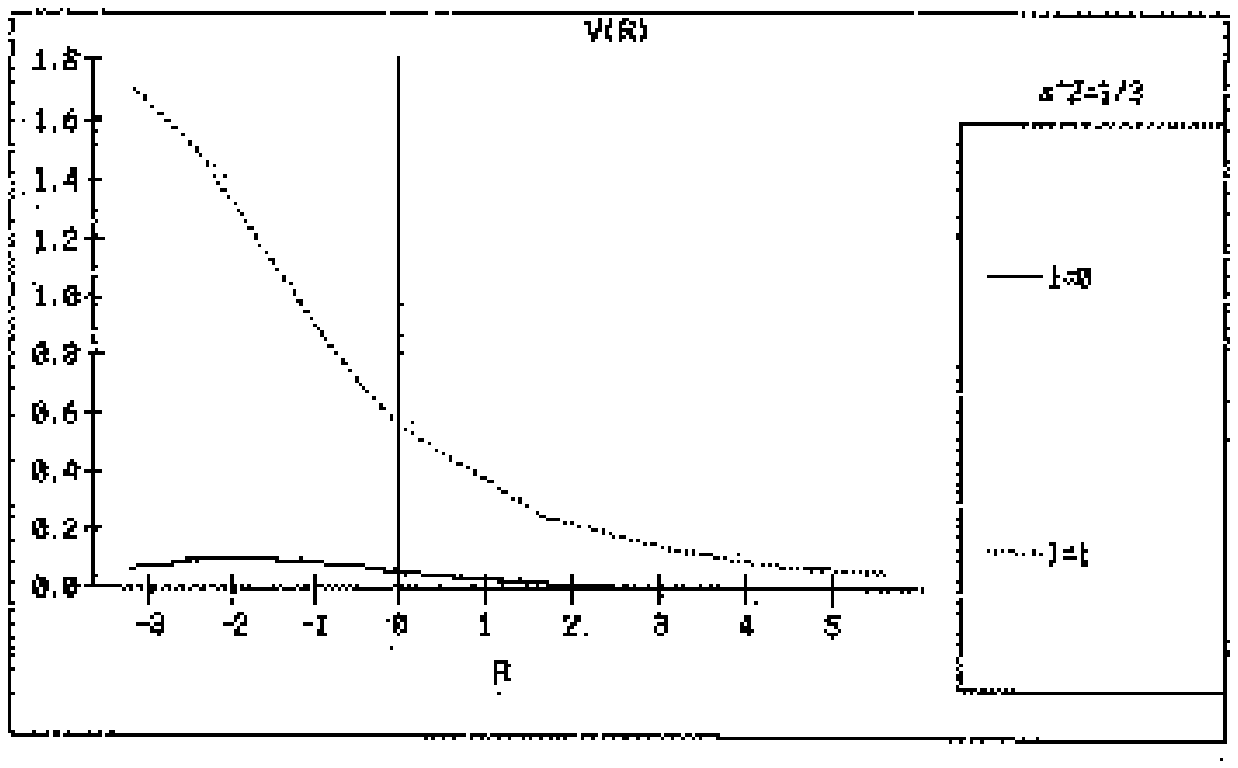}\\
(b)\\
\includegraphics[width=5cm]{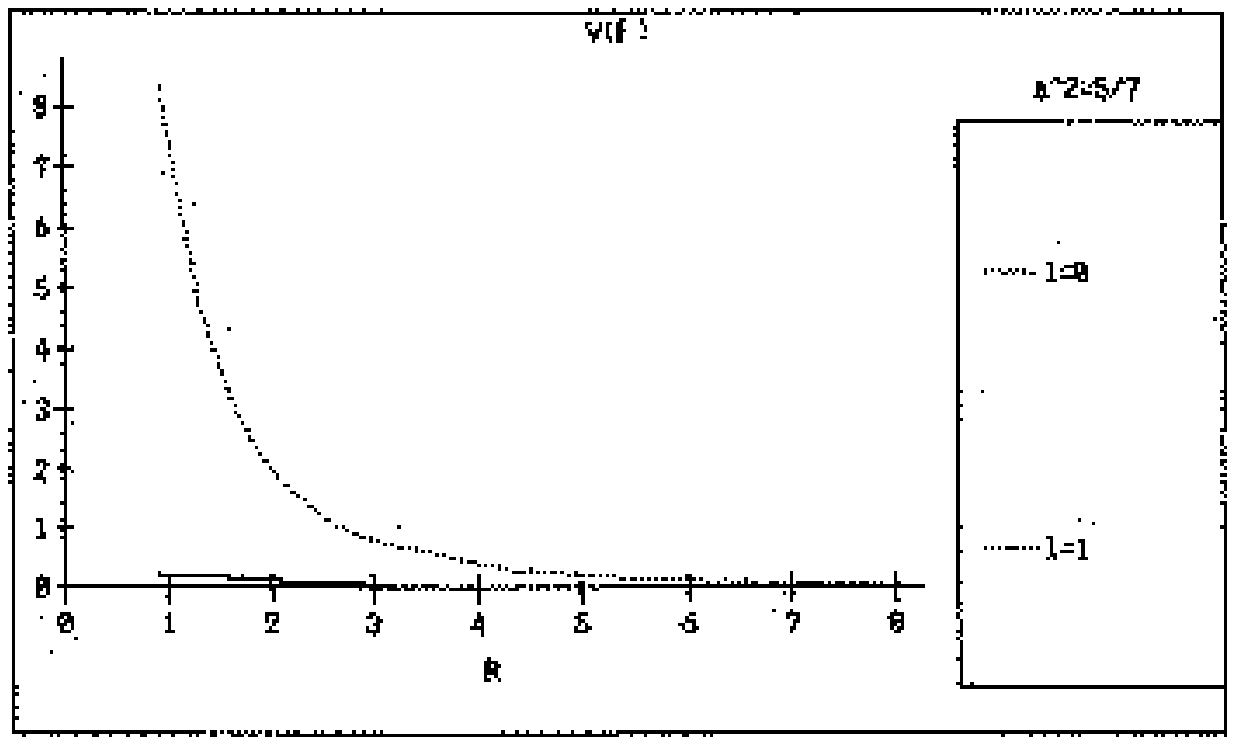}\\
(c)\\
\includegraphics[width=5cm]{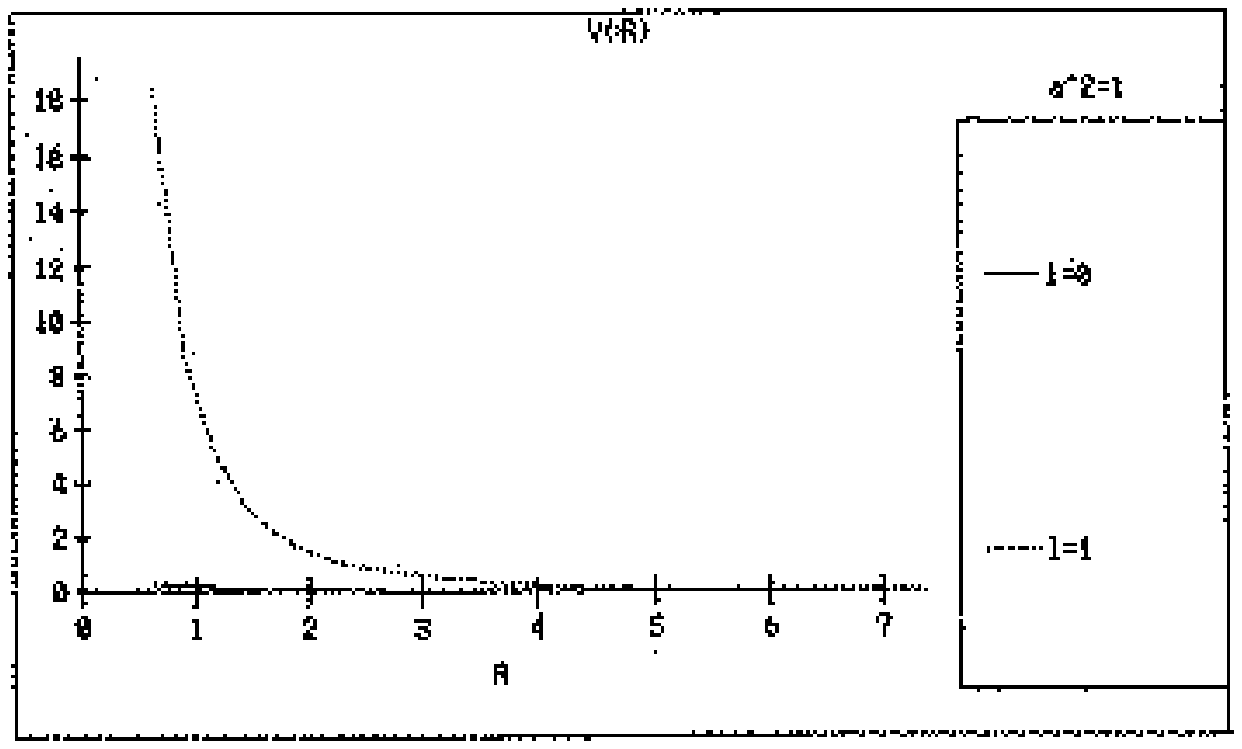}\\
(d)\\
\includegraphics[width=5cm]{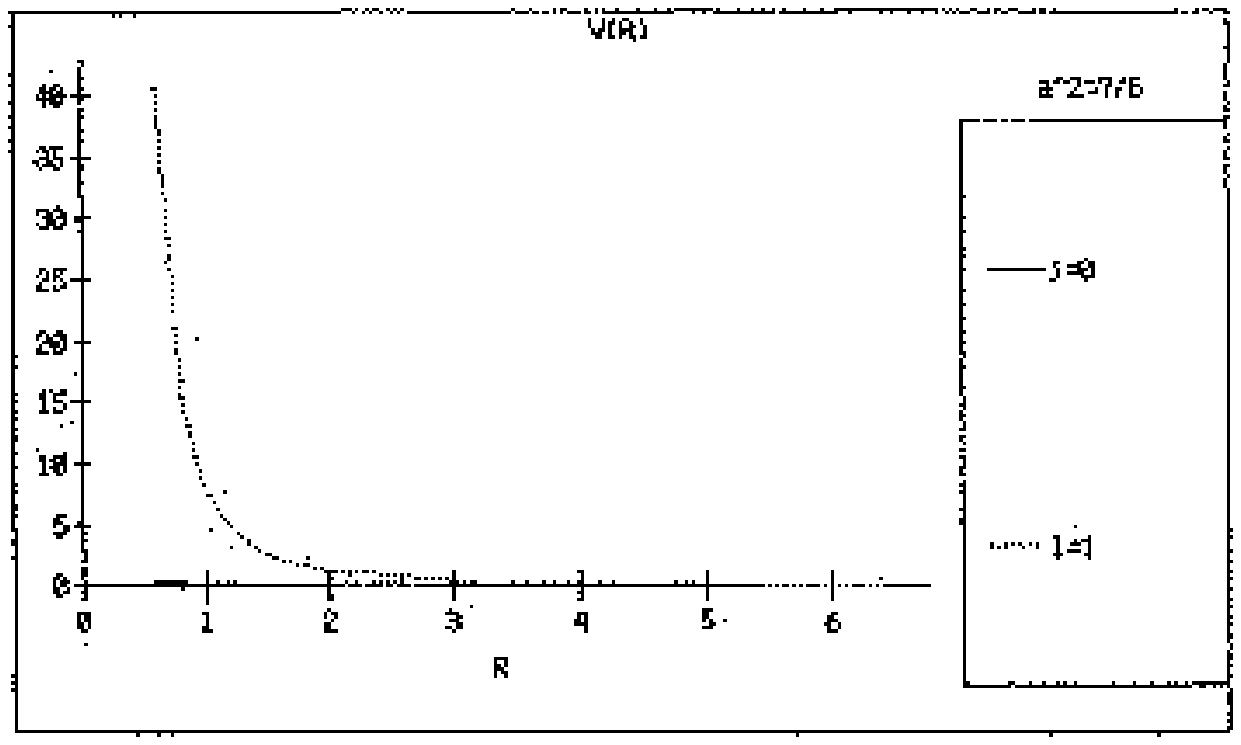}\\
(e)
\caption{The potential $V(R)$ is plotted against $R$, for several cases.
(a) $a^2=0$, (b) $a^2=1/3$, (c)	$a^2=5/7$, (d) $a^2=1$ and (e)
$a^2=7/5$. The $l=0$ and $l=1$ cases are shown in each figure. Here, $R$
is normalized by $r_+$ and $V(R)$ is normalized by $(r_+)^{-2}$, with
$r_+=(1+a^2)M$.}
\label{f1}
\end{center}
\end{figure}

\begin{figure}[ht]
\begin{center}
\includegraphics[width=5cm]{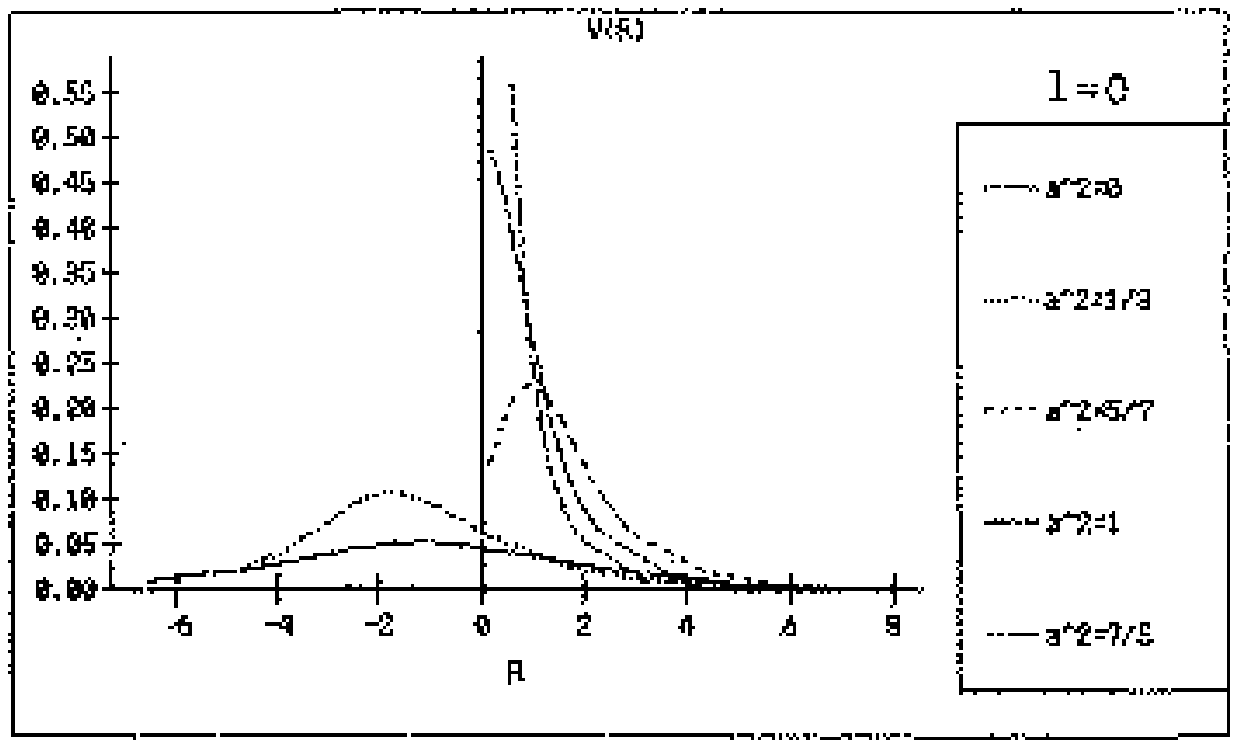}
\caption{The potential $V(R)$ for $l=0$ is plotted against $R$ in a
graph. The normalization is the same as in Fig.~1.}
\label{f2}
\end{center}
\end{figure}

The potentials for some peculiar values of $a^2$ are plotted in
Fig.~1(a-e) and Fig.~2. There are three cases which must be
distinguished. The qualitative feature of the potential for $a^2<1/3$ is
similar to that of the potential for $a^2=0$, which has been shown in
Fig.~1 of Ref.~\cite{14}. The case for $a^2=1/3$ is a special case,
where a barrier of an infinite height exists for $l\ne 0$.

For $1/3<a^2<1$, there is a similar barrier for $l\ne 0$, though $R$
ranges from $0$ to $+\infty$. The case with $a^2=1$ is another critical
case. For $1<a^2<3$, the infinite barrier at $R=0$ exists for all $l$.

The absorption of the wave into the origin is possible only if
$a^2\le 1/3$. The capture may occur dominantly in $l=0$ mode, as is
seen from the figures. The qualitative feature of capture cross-section
for the case with $a^2<1/3$ is similar to that for the case with
$a^2=0$, which has been studied intensively by Traschen and Ferrell
\cite{14}. We thus only make a few remarks here.

In a classical picture, a particle can go beyond a potential barrier
only if the amount of its energy is larger than the potential maximum.
This situation is realized when $q^2>V_{max}$, where $V_{max}$ is the
maximum height of the potential $V$. Semiclassical description is known
to be valid for a large angular momentum, or equivalently, a large impact
parameter. For large $l$, we obtain
\begin{equation}
V_{max}\sim\frac{4l(l+1)}{(1-3a^2)^2M^2}
\left(\frac{1-3a^2}{3-a^2}\right)^{(3-a^2)/(1+a^2)}\,.	
\label{4.14}
\end{equation}
On the other hand, the angular momentum is written in the semiclassical
region as $\sim\hbar qb\sim\hbar\sqrt{l(l+1)}$. Therefore the
capture by the black hole occurs when
\begin{equation}
b\sim\frac{\sqrt{l(l+1)}}{q}<\frac{1-3a^2}{2}
\left(\frac{3-a^2}{1-3a^2}\right)^{(3-a^2)/\{2(1+a^2)\}}M\,.	
\label{4.15}
\end{equation}
This condition agrees with the classical result, (\ref{3.11}). We thus
find that the semiclassical treatment yields the same result obtained by
the classical description.

For $a^2>1/3$, all waves are scattered back to infinity. This behavior
corre-sponds to the classical picture. Since our analysis is originally
based on the low-energy assumption, the partial-wave analysis on the
scattering problem may be most appropriate.

In the case with $a^2>1/3$, the asymptotic behavior of the partial wave
at infinity is expressed by using spherical Bessel functions:
\begin{eqnarray}
\frac{1}{R}\chi_l(R)&\sim& a_l j_l(qR)+b_l n_l(qR)\nonumber \\
&\sim&c_l\frac{1}{qR}\cos\left(qR-
\frac{l+1}{2}\pi+\delta_l\right)\quad\mbox{for }
R\rightarrow\infty\,,
\label{4.16}
\end{eqnarray}
where $\delta_l$ is called a phase shift and $a_l=c_l\cos\delta_l$,
$b_l=-c_l\sin\delta_l$.

The scattering amplitudes for partial waves are constructed from the
phase shift, i.e., \cite{27}
\begin{equation}
f_l(\Theta)=\frac{2l+1}{2iq}(e^{2i\delta_l}-1) P_l(\cos\Theta)\,,
\label{4.17}
\end{equation}
where $P_l(x)$ is a Legendre polynomial. The differential cross-section
$d\sigma/d\Omega$ is obtained from the amplitude
\begin{equation}
\frac{d\sigma}{d\Omega}=|f(\Theta)|^2\,,
\label{4.18}
\end{equation}
with
\begin{equation}
f(\Theta)=\sum_{l=0}^\infty f_l(\Theta)\,.
\label{4.19}
\end{equation}

Returning to the wave equation, we can find an equation by using the
constancy of the Wronskian. That is
\begin{equation}
\tan\delta_l=-q\int_0^\infty dR\,R\left[V(R)-\frac{l(l+1)}{R^2}
\right]j_l(qR)\frac{1}{a_l}\chi_l(qR)\,.
\label{4.20}
\end{equation}

First we consider small $q$. Then, $\chi_l(qR)$ can be approximated by
$a_lR j_l(qR)$. Consequently we obtain the result of the first Born
approximation:
\begin{equation}
\tan\delta_l=-q\int_0^\infty dR\,R^2\left[V(R)-\frac{l(l+1)}{R^2}
\right][j_l(qR)]^2\,.
\label{4.21}
\end{equation}

The most dominant contribution comes from the $S$ wave. For example, we
estimate the phase shift of the $S$ wave in the case of $a^2=1$.
Since $V(0)=1/(2r_+^2)$ and the half-width is roughly $r_+$ (as is seen
from Fig.~2), we find
\begin{equation}
\delta_0\sim\frac{-qr_+}{6}\qquad (a^2=1)\,,
\label{4.22}
\end{equation}
where $r_+=(1+a^2)M$. For higher modes, $\delta_l\sim-(qr_+)^{2l+1}$. The
total cross-section for the $S$ wave is roughly $4\pi r_+^2$, or rather
small.

Next, we examine the case with large $q$. The low velocity and large $q$
can be realized if $\hbar\ll 1$, because $v\sim\hbar q/\mu$. Hence, we
consider the semiclassical treatment to this end. The semiclassical
approximation is valid for the scattering problem if the variation of
the wavelength is smooth. We find the phase shift in this prescription:
\cite{27}
\begin{equation}
\delta_l=-qR_0+\int_{R_0}^\infty dR\,\{
[q^2-V(R)-1/(4R^2)]^{1/2}-q\}+\frac{2l+1}{4}\pi\,,
\label{4.23}
\end{equation}
where $R_0$ satisfies $V(R_0)=q^2$.

For $a^2>1/3$, the angular-momentum barrier at $R=0$ is dominant for
$l\ne 0$. In this case, we can use (\ref{4.12}) and obtain the following:
\begin{equation}
\delta_l=-\frac{\pi}{2}\left(\frac{3-a^2}{3a^2-1}\right)
\left[l+\frac{1}{2}-\frac{5a^2+1}{16(1+a^2)}
\left(l+\frac{1}{2}\right)^{-1}\right]
\,,
\label{4.24}
\end{equation}
which is valid for large $q$.

Again, we examine the $a^2=1$ case. Let us try to evaluate the total
cross-section including higher wave modes in this case. The total
cross-section is generally given by: \cite{27}
\begin{equation}
\sigma=\frac{4\pi}{q^2}\sum_l (2l+1) \sin^2\delta_l\,.
\label{4.25}
\end{equation}
Now, $\delta_l$ is $\simeq-(\pi/2)(l+1/2)$, and then we get
\begin{equation}
\sigma=\frac{2\pi}{q^2}\sum_{l=0}^{l_{max}} (2l+1)=\frac{2\pi}{q^2}
(l_{max}+1)^2\qquad	(a^2=1) .	
\label{4.26}
\end{equation}
If we regard $l_{max}$ as $\simeq qb$, the total cross-section diverges
owing to the contribution from $qb\gg 1$, or equivalently, at the small
deflection angle. This feature looks similar to the Coulomb scattering.
This fact agrees well with the classical analysis for $a^2=1$.

\section{Conclusion}
In this paper, we have examined the classical and the quantum scattering
problem of two maximally charged dilaton black holes. The system is
governed by a control parameter, a, which determines the strength of
coupling between a dilaton field and Maxwell field. We have found that
the nature of the velocity-dependent force at the lowest order is
sensitive to the dilaton coupling.

Therefore, the scattering of two dilaton black holes changes its manner
as $a^2$ takes a different value. We have studied the two-body scattering
by classical and quantum-mechanical analyses making use of the metric on
the moduli space.

The small-particle (wave) approximation is found to reproduce the same
result obtained by the analysis on the moduli space, both in classical
and quantum cases. 
We have also found a critical value for $a^2$ in the
scattering problem in this paper: that is $a^2=1/3$. Below this value,
the coalescence (or absorption) of black holes occurs for a sufficiently
small impact parameter. On the other hand, above the critical value, it
is impossible for two black holes to merge into one. This behavior can
be illustrated by both classical and quantum analyses on the moduli
space.

We have mainly examined the case where one of the black holes is much
more massive than the other (i.e., in the limit of infinitesimal reduced
mass of the system). We must note, however, that the nature of the
two-body problem at low energy is
expected not to be sensitive to the reduced mass. For example, the
moduli space  metric is independent of the reduced mass of the system
for the case with $a^2=1/3$  and $a^2=1$. Considering the geometry of
the moduli space, we can safely say that the characteristics of the
scattering are unchanged in quality by the finite reduced mass,
especially when $a^2$ takes a finite value (but $<3$).

We have not yet understood all the aspects of many-body forces in the
multi-maximally-charged-black-hole system. We must continue to make
every effort to solve the few-body problem.

The analysis on the moduli space can be generalized to the many-body
problems of black strings, black membranes, etc.~\cite{28}



\begin{thebibliography}{99}
\bibitem{1} N.~S.~Manton, Phys. Lett. \textbf{110} (1982) 54.
\bibitem{2} M.~Atiyah and N.~J.~Hitchin, \textit{The Geometry and
Dynamics of Magnetic Monopoles} (Princeton Univ. Press, 1988); Phys.
Lett. \textbf{A107} (1985) 21.
\bibitem{3} G.~W.~Gibbons and N.~S.~Manton, Nucl. Phys. \textbf{B274}
(1986) 183.
\bibitem{4}	B.~J.~Schroers, Nucl. Phys. \textbf{367} (1991) 177.
\bibitem{5} M.~P.~Wojtkowski, Bull. Am. Math. Soc. \textbf{18} (1988)
179.
\bibitem{6} M.~Temple-Raston, Phys. Lett. \textbf{206} (1988) 503; Nucl.
Phys. \textbf{B313} (1989) 447; Phys. Lett. \textbf{B213} (1988) 168.
\bibitem{7} P.~J.~Ruback, Nucl. Phys. \textbf{B296} (1988) 660.
\bibitem{8} E.~P.~S.~Shellerd and P.~J.~Ruback, Phys. Lett.
\textbf{B209} (1988) 262.
\bibitem{9} E.~Myers, C.~Rebbi and R.~Strilka., Phys. Rev. \textbf{D45}
 (1992) 1355.
\bibitem{10} T.~M.~Samols, Phys. Lett. \textbf{B244} (1990) 285.
\bibitem{11} T.~M.~Samols, Commun. Math. Phys. \textbf{145} (1992) 149.
\bibitem{12} G.~W.~Gibbons and P.~J.~Ruback, Phys. Rev. Lett.
\textbf{57} (1986) 1492.
\bibitem{13} R.~C.~Ferrell and D.~M.~Eardley, Phys. Rev. Lett.
\textbf{59} (1987) 1617.
\bibitem{14} J.~Traschen and R.~Ferrell, Phys. Rev. \textbf{D45}
(1992) 2628.
\bibitem{15} G.~W.~Gibbons and K.~Maeda, Nucl. Phys. \textbf{B298}
(1988) 741.
\bibitem{16} D.~Garfinkle, G.~Horowitz and A.~Strominger, Phys. Rev.
\textbf{D43} (1991), 3140; \textit{ibid.} \textbf{D45} (1992) 3888 (E).
\bibitem{17} J.~Preskill, P.~Schwarz, A.~Shapere, S.~Trivedi and
F.~Wilczek, Mod. Phys. Lett. \textbf{A6} (1991) 2353.
\bibitem{18} C.~F.~E.~Holzhey and F.~Wilczek, Nucl. Phys. \textbf{B380}
 (1992) 447.
\bibitem{19} K.~Shiraishi, Mod. Phys. Lett. \textbf{A7} (1992) 3449.
\bibitem{20} K. Shiraishi, Phys. Lett. \textbf{A166} (1992) 298.
\bibitem{21} J.~H.~Horne and G.~T.~Horowitz, Phys. Rev. \textbf{D46}
 (1992) 1340.
\bibitem{22} K.~Shiraishi, 
J. Math. Phys. \textbf{34} (1993) 1480.
\bibitem{23} K.~Shiraishi, 
Nucl. Phys. \textbf{B402} (1993) 399.
\bibitem{24} A.~Papapetrou, Proc. R. Irish Acad. \textbf{A51}
(1947) 191; S.~D.~Majumdar, Phys. Rev. \textbf{72} (1947) 930.
\bibitem{25} L.~D.~Landau and E.~M.~Lifschitz, \textit{Classical Theory
of Fields} (Pergamon, 1975), 4th edition.
\bibitem{26} L.~Infeld and J.~Plebanski, \textit{Motion and Relativity}
(Pergamon, 1960).
\bibitem{27} F.~Constantinescu and E.~Magyari, \textit{Problems in
Quantum Mechanics} (Pergamon, 1971).
\bibitem{28} A.~Dabholkar et al., Nucl. Phys. \textbf{B340} (1990) 33;
A.~Strominger, \textit{ibid.} \textbf{B343} (1990) 167; E.~Copeland,
D.~Haws and M.~Hindmarsh, Phys. Rev. \textbf{D42} (1990) 726; M.~J.~Duff
and J.~X.~Lu, Phys. Rev. Lett. \textbf{66} (1991) 1402; Nucl. Phys.
\textbf{B354}(1991) 129, 141 ; \textbf{B357} (1991) 534; Phys. Lett.
\textbf{B273} (1991) 409; M.~J.~Duff and K.~S.~Stelle, \textit{ibid.}
\textbf{B253} (1991) 113; R.~Guven, \textit{ibid.} \textbf{B276}
(1992) 49; R.~R.~Khuri, \textit{ibid.} \textbf{B259} (1991) 261;
C.~G.~Callan., Jr. and R.~R.~Khuri, \textit{ibid.} \textbf{B261}
(1991) 363; C.~G.~Callan, Jr., J.~A.~Harvey and A.~Strominger, Nucl.
Phys. \textbf{B359} (1991) 611; \textbf{B367} (1991) 60; G.~T.~Horowitz
and A.~Strominger, \textit{ibid.} \textbf{B360} (1991) 197; J.~A.~Harvey
and A.~Strominger, Phys. Rev. Lett. \textbf{66} (1991) 549;
S.~B.~Giddings and A.~Strominger, \textit{ibid.} \textbf{67}
(1991) 2930; D.~Brill and G.~T.~Horowitz, Phys. Lett. \textbf{B262}
(1991) 437; I. Martin and A. Restuccia, \textit{ibid.} \textbf{B271}
(1991) 361; J.~Horne, G.~Horowitz and A.~Steif, Phys. Rev. Lett.
\textbf{68} (1992) 568; R.~Guven, Phys. Lett. \textbf{B276} (1992) 49;
J.~H.~Horne and G.~T.~Horowitz, 
Nucl. Phys. \textbf{B368} (1992) 444;
J.~A.~Harvey and J.~Liu,
Phys. Lett. \textbf{B268} (1991) 40.
\end{thebibliography}
\end{document}